\documentclass[english,twoside,a4paper,10pt]{article}

\usepackage[latin1]{inputenc}
\usepackage[T1]{fontenc}
\usepackage{amsmath}
\usepackage{amsfonts}
\usepackage{graphicx}
\usepackage{a4wide}
\usepackage{amssymb}
\usepackage{fancyhdr}
\usepackage{mathrsfs}
\usepackage[toc,page]{appendix}

\begin{document}

\title{\bf Noether symmetry in  $F(T)$ gravity  with  $f$-essence}
\author{ 
Kairat Myrzakulov\footnote{e-mail address: kairatmyrzakul@gmail.com}, \,   Pyotr Tsyba\footnote{e-mail address: petr.tsyba@yahoo.com} \,  and  
Ratbay Myrzakulov\footnote{e-mail address: rmyrzakulov@gmail.com
}\\
\\
\begin{small} 
Eurasian International Center for Theoretical Physics, Eurasian National University, Astana, Kazakhstan
\end{small}\\
}

\date{}

\maketitle


\begin{abstract}
In $F(T)$  gravity theory, a  Friedman-Robertson-Walker cosmological model with  $f$-essence where fermion field is non-minimally coupled with the gravitational field is studied.  Using the Noether symmetry approach  the possible forms of $F(T)$ gravity and the  non-canonical fermionic lagrangian $K$ are determined. Cosmological solutions of the condered models  describing the accelerated and decelerated  periods of the    universe are found. 
\end{abstract}



\tableofcontents
\section{Introduction}

A new challenge for modern cosmology is to build some theoretical models to explain the dynamics of the large scale Universe. Different cosmological data supports the idea of an accelerating Universe  \cite{Perlmutter, Riess}. There are two approaches as attempts toresolve this problem:
One is to keep general relativity (GR) and add some exotic matter field(s), called dark energy (DE) \cite{Peebles, Copeland, Amendola}. The most viable candidates for the role of dark energy are some kinds of the scalar, fermion or tachyon fields as quintessence \cite{Okabe, Tsujikawa, Khurshudyan}, phantom energy \cite{Caldwell, Odintsov}, $k$-essence \cite{Mukhanov1, Mukhanov2}, $f$-essence \cite{Jamil:2011mc}-\cite{1107.1008}
 and $g$-essence \cite{Kulnazarov:2010an}-\cite{1012.5690}. However, this approach leads to some additional issues, such as the existence of negative entropy, future singularity energy condition violation and etc. Furthermore in this approach, all types of DEs, violate energy conditions as fundamental blocks of the GR. The second approach is called modified theories of gravity, and originally proposed by Buchdahl in  \cite{Buchdahl:1983zz},
when he generlized GR to a nonlinear form of Lagrangian 
 with an Ricci scalar $R$ on $F(R)$ gravity (see  \cite{Nojiri, Elizalde} for a revisit).
 Also, it was recently proposed another new modified theory of gravity so-called $F(T)$ gravity \cite{Bengochea:2008gz}
-\cite{1511.03655} , which is a generalization teleparallel gravity proposed by Einstein. In this theory is used the Weitzenb$\ddot{o}$ck connection, then as in GR used the Levi-Civita connection.
Next generalization of $F(R)$ and $F(T)$ gravity theory is $F(R,T)$ gravity , withdiffrent meanings for $T$, the one for trace of the energy-momentum tensor of  and its higher derivative extensions \cite{Harko:2011kv}-\cite{Houndjo:2016rlg}.\par
Symmetry is an essential feature to build any theory of particle physics and gravity. If we find the point like Lagrangian, we can study the family of symmetries which are corresonding to the conserved quantities. These types of the symmetries are called Noether symmetry. The conserved quantity which is associated to the symmetry generator is called as Noether charge, in correspondence to the same terminology in particle physics and gauge theories.

In recent years Noether symmetry successfully used for the construction of various cosmological models \cite{Capozziello:2012hm}-\cite{noether2}. Noether symmetry has been used to investigate cosmology in phantom quintessence cosmology, for non-minimally coupled fermion fields,for boson and fermion field, in $F(R)$ cosmology,  in teleparallel gravity,  in $F(T)$ gravity,  in $F(R,T)$ gravity and Noether symmetry in quantum cosmology \cite{Souza}-\cite{1311.2173}.

In this paper, we consider Noether symmetry for  $F(T)$ gravity  with  f-essence, as a framework to explain the present accelerated expansion of the Universe. The matter component is assumed to be a generic function of the kinetic energy of the fermion field. To find the explicit form of the function $F(T)$, we utilize generalized Noether theorem and use generalized vector fields as infinitesimal generators of the symmetries for the corresponding Lagrangian. We study the cosmological consequences of the obtained results.

This paper is organized as follows: in Sec. 2, we review the basics of $f(T)$ gravity. In Sec. 3, we derive the equations of motions for $F(T)$ gravity with $f$-essence with non-minimally coupled to the gravitational field from a point-like Lagrangian in a spatially flat Friedman-Robertson-Walker metric (FRW). In  Sec. 4, we look for the existence of Noether symmetry for the point like Lagrangian is the subject, where the possible forms of the coupling and of the potential density are determined. In Sec. 5, the field equations are solved for couplings and potential densities found in the previous sections and the cosmological solutions are investigated. The final remarks and conclusions are the subject of the last section.

We adopted the signature as $\left(+,-,-,-\right)$ and natural units $8\pi G = c = \hbar = 1$.


\section{$F(T)$ gravity}

The action for $F(T)$ gravity is given by 
\begin{equation}
 \label{action} 
S  = \int d^4  x e \left[F(T)+L_{m}\right],
\end{equation}
where $e=\det\left(e^{i}_{\mu}\right)=\sqrt{-g}$, $T$ is the torsion scalar and $L_{m}$ stands for the matter Lagrangian. 

The torsion scalar  is defined by the expression,
\begin{equation}
\label{} 
T = S_{\rho}^{\mu \nu} T^{\rho}_{\mu \nu},
\end{equation}
where $S_{\rho}^{\mu \nu}$ is given as follows
\begin{equation}
\label{} 
S_{\rho}^{\mu \nu} = \frac{1}{2} \left(K^{\mu \nu}_{\rho} + \delta_{\rho}^{\mu} T^{\theta \nu}_{\theta} - \delta_{\rho}^{\nu} T^{\theta \mu}_{\theta}\right),
\end{equation}
and torsion tensor $T^{\rho}_{\mu \nu}$ is given as
\begin{equation}
\label{} 
T^{\rho}_{\mu \nu} = \Gamma^{\rho}_{\nu \mu} - \Gamma^{\rho}_{\mu \nu} = e_{i}^{\rho} \left(\partial_{\mu} e_{\nu}^{i} - \partial_{\nu} e_{\mu}^{i}\right).
\end{equation}
Here $e_{\mu}^{i}$ are the components of the non-trivial tetrad  field $e_{i}$ in the coordinate basis. There is a wide extra choiceto find a good  tetrad basis. The reason backs to the  following identity regsrding metric in vielbein formalism:
\begin{equation}
\label{}
g_{\mu \nu} = \eta_{ij} e_{\mu}^{i} e_{nu}^{j},
\end{equation}
where $\eta_{ij} = diag(1,-1,-1,-1)$ is the Minkowski metric for the tangent space. Consequently, with a given metric, there exist infinite set of basis tetrad fields $e_{\mu}^{i}$ all can satisfy the following properties:
\begin{equation}
\label{}
e_{\mu}^{i} e_{j}^{\mu} = \delta_{j}^{i}, \ \ e_{\mu}^{i} e_{i}^{\nu} = \delta_{\mu}^{\nu}.
\end{equation}
The procedure for evaluating the tetrad field has been studied in literature by many authors but so far, there is not fixed manifest to find a good strategy to exclude bad tetrads.  Notice that the Latin alphabets $\left(i,j,\mu, \nu = 0,1,2,3\right)$ will be used to denote the tangent space indices and the Greek alphabets $\left(\mu, \nu, ... = 0,1,2,3\right)$ to denote the spacetime indices. The contorsion tensor $K^{\mu \nu}_{\rho}$ is defined as
\begin{equation}
\label{}
K^{\mu \nu}_{\rho} = - \frac{1}{2} \left(T^{\mu \nu}_{\rho} - T^{\nu \mu}_{\rho} - T_{\rho}^{\mu \nu}\right),
\end{equation}
which is equal to the difference between a pair of Weitzenb$\ddot{o}$ck connections. The variation of action (\ref{action}) with respect to the vierbein field leads to the following field equations,
\begin{equation}
\label{}
\left[e^{-1} \partial_{\mu} \left(e S_{i}^{\mu}\nu\right) - e_{i}^{\lambda} T^{\rho}_{\mu \lambda} S_{\rho}^{\nu \mu}\right] F_{T} + S_{i}^{\mu \nu} \partial_{\mu} (T) F_{TT} + \frac{1}{4} e_{i}^{\nu} F = \frac{1}{2} k^2 H_{i}^{\rho} T_{\rho}^{\nu}.
\end{equation}
Here $T_{\mu \nu}$ is the energy-momentum tensor given as
\begin{equation}
\label{}
T_{\rho}^{\nu} = diag\left(\rho_{m}, -p, -p, -p\right),
\end{equation}
where $\rho_{m}$ and $p_{m}$ are the density and presssure of matter fields inside the Universe.

\section{$F(T)$ gravity with $f$-essence}

In this section, we derived the equations of motions for $F(T)$ gravity with $f$-essence with non-minimally coupling to the gravitational background. 
The gravitational action for $f$-essence is given by the following expression:
\begin{equation}
\label{}
S = \int d^4 x e \left[F(T) + 2 K (Y, u)\right],
\end{equation}
where $K$ is some function of its arguments, $u=\bar{\psi} \psi $, $\psi = \left(\psi_{0}, \psi_{1}, \psi_{2}, \psi_{3}\right)^{T}$ is a fermionic function and $\bar{\psi}=\psi^{+} \gamma^{0}$ denotes its adjoint function. Furthermore, the kinetic part is defined by the following:
\begin{equation}
\label{}
Y = 0.5 i \left[\bar{\psi} \Gamma^{\mu} D_{\mu} \psi - \left(D_{\mu} \bar{\psi}\right) \Gamma^{\mu} \psi \right],
\end{equation}
Here the differential operator $D_{\mu}$ is the covariant derivative and $\Gamma^{\mu} = e^{\mu}_{a} \gamma^{a}$.

We will consider here the simplest homogeneous and isotropic cosmological model,  with the following metric:

\begin{eqnarray} \label{FRW}
& & ds^2  = dt^2 -  a^2(t)\big[ dx^2 + dy^2 + dz^2\big],
\end{eqnarray}
where $a(t)$ is the scale factor of the Universe. For this metric, the good vierbein basis found to be $(e^\mu_a)=diag(1,1/a,1/a,1/a)$ and $(e^a_\mu)=diag(1,a,a,a)$.

The Dirac matrices of curved spacetime $\Gamma^\mu$ are
\begin{eqnarray}
\Gamma^0=\gamma^0,\ \  \Gamma^j=a^{-1}\gamma^j,\ \  \Gamma^5=-i\sqrt{-g}\Gamma^0\Gamma^1\Gamma^2\Gamma^3=\gamma^5,\ \  \Gamma_0=\gamma^0,\ \  \Gamma_j=a\gamma^j (i=1,2,3).
\end{eqnarray}

Hence we get 
\begin{eqnarray}
\Omega_0=0,\ \ \Omega_j=\frac{1}{2}\dot{a}\gamma^j\gamma^0
\end{eqnarray}
The kinetic part is written as the following:
\begin{eqnarray}
Y=\frac{1}{2}i\left(\bar{\psi}\gamma^0\dot{\psi}-\dot{\bar{\psi}}\gamma^0\psi\right).
\end{eqnarray}
Finally, we note that the gamma matrices are given as follow:
 \begin{equation}
\gamma^0 = \begin{pmatrix} I & 0 \\ 0 & -I \end{pmatrix}, \quad\gamma^k = \begin{pmatrix} 0 & \sigma^k \\ -\sigma^k & 0 \end{pmatrix},\quad \gamma^5 = \begin{pmatrix} 0 & I \\ I & 0 \end{pmatrix},
\end{equation}
here $I=diag (1,1)$ and the $\sigma^k$ are Pauli $SU(2)$ matrices having the following form
 \begin{equation}
\sigma^1 = \begin{pmatrix} 0 & 1 \\ 1 & 0 \end{pmatrix},\quad \sigma^2 = \begin{pmatrix} 0 & -i \\ i & 0 \end{pmatrix},\quad \sigma^3 = \begin{pmatrix} 1 & 0 \\ 0 & -1 \end{pmatrix}.
\end{equation}

In fact selecting suitable Lagrange multipliers and integrating by parts to eliminate higher order derivatives, the Lagrangian ${\cal L}$ transforms to the  canonical form. In physical units, the action is
\begin{eqnarray}\label{s2}
 S =\int d^4x \left[a^3\left(h(u)F(T) - \lambda \left(T + 6 H^2 \right) + 2 K(Y,u)\right)\right],
\end{eqnarray}
where $u=\bar{\psi} \psi$ is the bilinear function. In order to determine $\lambda$, we have to vary the action with respect to $T$, that is
\begin{equation}
h \frac{d F}{d T} \delta T - \lambda \delta T =0,
\end{equation}
in which we obtain:
\begin{equation} 
\label{}
\lambda = h F_{T}.
\end{equation}
where $\lambda$ is a Lagrange multiplier. 

Therefore, the action (\ref{s2}) can be rewritten as
\begin{equation}
\label{}
 S =\int d^4x \left[a^3 \left(h F - h F_{T} \left(T + 6 H^2 \right) + 2 K\right)\right],
\end{equation}
and then the point-like Lagrangian reads
\begin{eqnarray}\label{L}
L = a^3 hF - a^3 h F_{T} T - 6 a \dot{a}^2 h F_{T} + 2 a^3 K.
\end{eqnarray}

It is well known that, for a dynamical system, the Euler-Lagrange equation is defined by the following:
\begin{eqnarray}
\label{EL}
\frac{\partial L}{\partial q_{i}} - \frac{d}{dt} \left(\frac{\partial L}{\partial \dot{q}_{i}}\right) =0,
\end{eqnarray}
where $q_{i}$ are the generalized coordinates of the configuration space $Q$, and in our case $q_{i} = a, \psi, \bar{\psi}$ and $T$. Substituting the Lagrangian (\ref{L}) into the Euler-Lagrange equation (\ref{EL}), we obtain
\begin{eqnarray}
\label{eqs}
F_{TT}\left(T + 6 H^2 \right)&=&0,\\
4 H F_{TT} \dot{T} +  \left( 6 H^2 + 4 \dot{H} - T + 4 H \frac{\dot{h}}{h} \right)F_{T} + F + \frac{2}{h} K&=&0,\\
K_{Y} \dot{\psi} + 0.5 \left(3 H K_{Y} + \dot{K}_{Y}\right) \psi - i K_{u}  \gamma^{0} \psi  - 0.5 i  \left(F-F_{T} T - 6 H^2 F_{T}\right) h_{u} \gamma^{0} \psi &=&0,\\
K_{Y} \dot{\bar{\psi}} + 0.5 \left(3 H K_{Y} + \dot{K}_{Y}\right) \bar{\psi} + i K_{u} \bar{\psi} \gamma^{0} + 0.5 i  \left(F-F_{T} T - 6 H^2 F_{T}\right) h_{u} \bar{\psi} \gamma^{0}&=&0,
\end{eqnarray}
were $H=\frac{\dot{a}}{a}$ denotes the Hubble parameter. From equations (\ref{eqs}) one can be seen $T=-6H^2$, which corresponds to the basic consideration of model.

We also consider the energy condition 
\begin{equation}
\label{E}
E=\frac{\partial L}{\partial \dot{a}} \dot{a} + \frac{\partial L}{\partial \dot{T}} \dot{T} + \frac{\partial L}{\partial \dot{\psi}} \dot{\psi} + \frac{\partial L}{\partial \dot{\bar{\psi}}}\dot{\bar{\psi}}   - L,
\end{equation}
Here the dot indicates the derivatives with respect to the cosmic time $t$. By combining the equations (22) and (28), we obtain
\begin{equation}
\label{}
 \left(F_{T} T - 6H^2 F_{T} - F\right) h + 2 \left( Y  K_{Y}- K \right)=0. 
\end{equation}
So finally, we have the following system of equations for our model:
\begin{eqnarray}
\label{eq1}
12H^2 F_{T} + F -\frac{2}{h} \rho_{f}&=&0,\\\label{eq2}
48 H^2 \dot{H} F_{TT} - 4 \left(3 H^2 + \dot{H}  + H \frac{\dot{h}}{h}\right) F_{T} - F  -\frac{2}{h} p_{f}&=&0,\\ \label{eq3}
K_{Y} \dot{\psi} + 0.5 \left(3 H K_{Y} + \dot{K}_{Y}\right) \psi - i K_{u}  \gamma^{0} \psi - 0.5 i  \left(F-F_{T} T - 6 H^2 F_{T}\right) h_{u} \gamma^{0} \psi &=&0,\\ \label{eq4}
K_{Y} \dot{\bar{\psi}} + 0.5 \left(3 H K_{Y} + \dot{K}_{Y}\right) \bar{\psi} + i K_{u} \bar{\psi} \gamma^{0} + 0.5 i  \left(F-F_{T} T - 6 H^2 F_{T}\right) h_{u} \bar{\psi} \gamma^{0}&=&0.
\end{eqnarray}
where
\begin{equation}
\label{}
 \rho_{f} =YK_{Y}  - K, \quad p_{f}=K 
\end{equation}  are  the energy density and pressure of the fermionic field. From the equations (\ref{eq1})-(\ref{eq4}), we see that these equations are nonlinear differential equations, respectively are difficult to find they solutions. To solve these equations need to find a form of the function $F(T), h(u)$ and $K(Y,u)$. In the next section we will use Noether symmetry approach for find these function. 

\section{The Noether Symmetries Approach}

The basic idea of 
Noether symmetry approach is that to find a class of symmetry generators ${\bf X}$ with those generatrs, the Lie derivative of the Lagrangian  vanishes, i.e. \begin{equation}\label{noether}
\mathcal{L}_{\bf X} = 0. 
\end{equation}  
Our plan is to look for a possible set of the Noether symmetries for our model in terms of the components of the spinor
field $\psi= (\psi_0, \psi_1, \psi_2, \psi_3)^T$ and its adjoint $\bar{\psi} =
({\psi_0}^\dagger, {\psi_1}^\dagger, -{\psi_2}^\dagger, -{\psi_3}^\dagger)$. The existence of Noether symmetry given by the equation (\ref{noether}) implies the existence of a vector field ${\bf X}$ such that
\begin{eqnarray} 
{\bf X}=\alpha \frac{\partial}{\partial a}+ \beta\frac{\partial}{\partial T}+{\dot \alpha} \frac{\partial}{\partial  \dot a}+  {\dot \beta}\frac{\partial}{\partial \dot T}+\sum^3_{j=0}\left(\eta_j  \frac{\partial}{\partial \psi_j}+{\dot \eta_j}  \frac{\partial}{\partial\dot \psi_j}+\chi_j  \frac{\partial}{\partial \psi^\dagger_j}+{\dot \chi_j}  \frac{\partial}{\partial \dot \psi^\dagger_j}\right)\,,
\label{ourX}
\end{eqnarray}
where $\alpha, \beta, \eta_i$ and $\chi_i$ are depend on $a, T, \psi_{i}$ and $\psi_{i}^\dagger$ and their derivatives are determined from the following equations 
\begin{eqnarray} 
{\dot \alpha}\,&=&\, \frac{\partial \alpha}{\partial a}{\dot a}+\frac{\partial \alpha}{\partial T}{\dot T}+ \sum^3_{i=0}\left(\frac{\partial\alpha }{\partial \psi_i}\dot{\psi_i}+\frac{\partial\alpha}{\partial \psi^\dagger_i}\dot{\psi^\dagger_i}\right),\\
{\dot \beta}\,&=&\, \frac{\partial \beta}{\partial a}{\dot a}+\frac{\partial \beta}{\partial T}{\dot T}+\sum^3_{i=0}\left( \frac{\partial\beta}{\partial \psi_i}\dot{\psi_i}+\frac{\partial\beta}{\partial \psi^\dagger_i}\dot{\psi^\dagger_i}\right),\\
{\dot \eta_{j}}\,&=&\, \frac{\partial \eta_{j}} {\partial a}{\dot a}+\frac{\partial \eta_{j}}{\partial T}{\dot T}+\sum^3_{i=0}\left( \frac{\partial\eta_{j} }{\partial \psi_i}\dot{\psi_i}+\frac{\partial\eta_{j}}{\partial \psi^\dagger_i}\dot{\psi^\dagger_i}\right),\\
{\dot \chi_{j}}\,&=&\, \frac{\partial \chi_{j}}{\partial a}{\dot a}+\frac{\partial \chi_{j}}{\partial T}{\dot T}+\sum^3_{i=0}\left( \frac{\partial \chi_{j}}{\partial \psi_i}\dot{\psi_i}+\frac{\partial \chi_{j}}{\partial \psi^\dagger_i}\dot{\psi^\dagger_i}\right),
\end{eqnarray}
 
The condition (\ref{noether}) when applied to the Lagrangian (\ref{L}) leads to an ploynomial expression of  $\dot{a}^2, \, \dot{T}^2, \, \dot{a}\dot{T}, \, \dot{a} \dot{\psi}^{\dagger}_{i}, \, \dot{a} \dot{\psi}_{i}, \, \dot{T} \dot{\psi}^{\dagger}_{i}, \, \dot{T} \dot{\psi}_{i}, \, \dot{\psi}^{\dagger}_{i} \dot{\psi}_{i}, \, \dot{a}, \, \dot{T}, \, \dot{\psi}^{\dagger}_{i}$ and $\dot{\psi}_{i}$. Puuting the coefficients of the above expressionset to the zero, one obtains the following system of coupled differential equations: 
\begin{eqnarray}
\label{}
\alpha F_{T} + \beta a F_{TT} + 2 a F_{T} \frac{\partial \alpha}{\partial a} + a F_{T} \frac{h_{u}}{h} \sum_{i=0}^{3} \left(\epsilon_{i} \eta_{i} \psi^{\dagger}_{i} + \epsilon_{i} \chi_{i} \psi_{i}\right)&=& 0, \\
3 \alpha \left(F-F_{T} T \right) - \beta a F_{TT} T + a \frac{h_{u}}{h} \left(F-F_{T} T\right) \sum_{i=0}^{3} \left(\epsilon_{i} \eta_{i} \psi^{\dagger}_{i} + \epsilon_{i} \chi_{i} \psi_{i}\right)&=&0, \\  3 \alpha \left(K - Y K_{Y}\right) + a K_{u}\sum_{i=0}^{3} \left(\epsilon_{i} \eta_{i} \psi^{\dagger}_{i} + \epsilon_{i} \chi_{i} \psi_{i}\right)&=&0, \\ F_{T} \frac{\partial \alpha}{\partial T} &=& 0, \\ F_{T} \frac{\partial \alpha}{\partial \psi_{i}} &=& 0, \\ F_{T} \frac{\partial \alpha}{\partial \psi^{\dagger}_{i}} &=&0, \\ \sum_{i=0}^{3} \left(\frac{\partial \eta_{i}}{\partial a} \psi^{\dagger}_{i} - \frac{\partial \chi_{i}}{\partial a} \psi_{i}\right)&=&0, \\ \sum_{i=0}^{3} \left(\frac{\partial \eta_{i}}{\partial T} \psi^{\dagger}_{i} - \frac{\partial \chi_{i}}{\partial T} \psi_{i}\right)&=&0, \\ 
3 \alpha \psi_{j}  +  a\eta_{i} - a\sum_{i=0}^{3} \left(\frac{\partial \eta_{i}}{\partial \psi^{\dagger}_{i} } \psi^{\dagger}_{i} - \frac{\partial \chi_{i}}{\partial \psi^{\dagger}_{i}} \psi_{i}\right)&=&0, \\  3 \alpha \psi^{\dagger}_{j} + a \chi_{j} + a\sum_{i=0}^{3} \left(\frac{\partial \eta_{i}}{\partial \psi_{j}} \psi^{\dagger}_{i} - \frac{\partial \chi_{i}}{\partial \psi_{j}} \psi_{i}\right) &=&0.
\end{eqnarray}
where $\epsilon_{i} = 1$ for $i=1,2$ and $\epsilon_{i} = -1$ for $i=3,4$. From equations (43)-(45) we can see that the function $\alpha$ dependent  only a function of $a$. We can rewrite equation (42) as
\begin{equation}
\frac{3 \alpha}{a K_{u}} \left(K - Y K_{Y} \right) = - \sum_{i=0}^{3} \left(\epsilon_{i} \eta_{i} \psi^{\dagger}_{i} + \epsilon_{i} \chi_{i} \psi_{i}\right).
\end{equation}

If we put this equation into equations (40)-(41), then the corresponding results are
\begin{equation}
\alpha F_{T} + \beta a F_{TT} + 2 a F_{T} \frac{\partial \alpha}{\partial a} -  \frac{3 \alpha F_{T} h_{u} }{h K_{u}} \left(K - Y K_{Y}\right) = 0,
\end{equation}
\begin{equation}
3 \alpha \left(F-F_{T} T \right) - \beta a F_{TT} T - \frac{3 \alpha h_{u}}{h K_{u}} \left(F-F_{T} T\right) \left(K - Y K_{Y}\right) = 0,
\end{equation}
The equation (52), we can rewrite as 
\begin{equation}
\beta a F_{TT} = \frac{3 \alpha}{T} \left(F-F_{T} T \right) \left[1 - \frac{ h_{u}}{h K_{u}} \left(K - Y K_{Y}\right)\right],
\end{equation}
Put this equation in equation (51) and after some algebraic calculations, we have the following equation:
\begin{equation}
\frac{a}{\alpha}  \frac{\partial \alpha}{\partial a} = 1 - \frac{3 F}{2 T F_{T}} \left[1 - \frac{ h_{u}}{h K_{u}} \left(K - Y K_{Y}\right)\right] = n,
\end{equation}
where $n$ is a constant. Then, we find generator $\alpha$ as
\begin{equation}
\alpha = \alpha_{0} a^{n}.
\end{equation}
where $\alpha_{0}$ is a constant of integration. Now from the equation (54) and considering that $h$ and $K$ depend only on a function $u$ and the function $F$  is dependent of $T$, we have the following equation
\begin{equation}
\frac{h_{u} }{h K_{u}} \left(K - Y K_{Y}\right) = 1 + \frac{2\left(n-1\right) T F_{T}}{3 F} = m,
\end{equation}
where $n$ is a constant. From this equation, we have the following pair of ordinary differential equations
\begin{equation}
2 \left(n-1\right) T F_{T} - 3 (m-1) F = 0,
\end{equation}
\begin{equation}
\frac{K - Y K_{Y}}{K_{u}} = m \frac{h}{h_{u}}.
\end{equation}

From the first equation we determine the form of the function $F$ and its derivatives as
\begin{equation}
F=C_{1} T^{\frac{3 \left(m-1\right)}{2 \left(n-1\right)}}.
\end{equation}

Also, from the equations (48), (49) and (50), we find the solutions for the generators $\eta_{j}$ and $\chi_{j}$ as 
\begin{equation}
\eta_{j} = - \left(\frac{3}{2} \alpha_{0} a^{n-1} + \epsilon_{j} \eta_{0}\right) \psi_{j},
\end{equation}
\begin{equation}
\chi_{j} = - \left(\frac{3}{2} \alpha_{0} a^{n-1} - \epsilon_{j} \eta_{0}\right) \psi^{\dagger}_{j}.
\end{equation}
Substituting these values into equation (50) and using the $\alpha$ given in (55), we have the following equation
\begin{equation}
\frac{K - Y K_{Y}}{K_{u}}  = u,
\end{equation}
or
\begin{equation}
K - Y K_{Y} - u K_{u} = 0,
\end{equation}

To solve this equation we need to make the change of variables
\begin{equation}
Y,u \rightarrow p = Y - \nu u, \ \frac{\partial}{\partial Y} = \frac{\partial}{\partial p}, \  \frac{\partial}{\partial u} = - \nu \frac{\partial}{\partial p}.
\end{equation}
where $\nu$ is an another constant. Then the partial differential equation (63) is transformed into an ordinary differential equation
\begin{equation}
K - p \frac{d K}{d p} = 0,
\end{equation}
Then, we find $K$ as
\begin{equation}
K = K_{0} \left(Y - \nu u\right)
\end{equation}
where $K_0$ is a integrable constant.

If you compare the equation (58) with equation (66) will be available the following expression
\begin{equation}
h = h_{0} u^{m}
\end{equation}

Finally, we put the equations (55), (59), (66) and (68) into (53), we are obtained
\begin{equation}
\beta = 2 \alpha_{0} \left(n-1\right) a^{n-1} T.
\end{equation}

Thus, we find the explicit non-zero solutions for the functions $\alpha, \, \beta, \, \eta_{i}, \, \chi_{i}, F, K$ and $h$. Therefore, for our model Noether symmetry vector $\bf{X}$ exists.

\section{Cosmological solutions}
In previous section we obtained the form of Noether symmetry for our model. Now we're ready to solve equations of motion and to find exact cosmological solutions. As a first step, 
weneed to substitute the solutions (59), (66) and (67)in  the field equations (32), (33). Then, we have 
\begin{equation}
\dot{u} + 3 \frac{\dot{a}}{a} u =0,
\end{equation}
so that 
\begin{equation}
u = \frac{u_{0}}{a^3}
\end{equation}
where $u_{0}$ is an integration constant. 

In order to determine the function dependence of the scale factor of the time $a(t)$, we need to substitute the values of functions $F(T), h(u)$ and $K(Y,u)$ obtained from equations (59), (66) and (68) into equation (30), then we have 
\begin{equation}
\dot{a} = a_{0} a^{n}.
\end{equation}

The solution of this differential eqation is given by:
\begin{equation}
a (t) = \left[a_{0} \left(1-n\right) \left(t-C_{2}\right)\right]^{\frac{1}{1-n}},
\end{equation}
where $C_{2}$ is a constant of integration, $n \neq 1$ and the constant $a_{0}$ is
\begin{equation}
a_{0} = \frac{1}{\sqrt{-6} u_{0}^{\frac{1}{3} \left(n-1\right)}} \left[\frac{2 K_{0} \nu \left(n-1\right)}{h_{0} C_{1} \left(n+2-3m\right)}\right]^{\frac{n-1}{3\left(m-1\right)}}.
\end{equation}
and we can see
\begin{equation}
u = u_{0} \left[a_{0} \left(1-n\right) \left(t-C_{2}\right)\right]^{- \frac{1}{1-n}}.
\end{equation}

We find that the Hubble parameter is
\begin{equation}
H=\frac{\dot{a}}{a} = \frac{1}{\left(1-n\right)\left(t-C_{2}\right)}.
\end{equation}

The energy density and pressure for this model are
\begin{equation}
\rho = \frac{3}{\left(1-n\right)^{2}\left(t-C_{2}\right)^{2}},
\end{equation}
\begin{equation}
p = - \frac{2n+1}{\left(1-n\right)^{2}\left(t-C_{2}\right)^{2}}.
\end{equation}

The equation of state parameter for our model can be define as
\begin{equation}
\omega = \frac{p}{\rho} = - \frac{1}{3} \left(1+2n\right).
\end{equation}
As was shown earlier for our model, the constant $n\neq1$. In our model, we consider a value $n>1$ then we have $\omega < -1$ that this phase is the phantom phase and if $n=0$ we have $\omega = - \frac{1}{3}$ is the quintessence phase.  

The decelaration parameter for the fermionic field define as
\begin{equation}
q = - \frac{\ddot{a} a }{\dot{a}^2} = - n,
\end{equation}
From this example we can see that for $n>0$ of our universe can be the the accelerated, for $n<0$ decelerating expansion. 

When $n=-\frac{1}{2}$, we can see
\begin{equation}
\rho_{tot} = \frac{4}{3 \left(t-C_{2}\right)^{2}}, \  p=0.
\end{equation} 
From this example we have a standard pressureless matter field. As a vital fact, we conclude that the fermionic field behaves as both the phantom and quintessence phase of the accelerating expanding universe.

\section{Conclusions}
\label{cinque}  

In this paper we have considered the Noether symmetry approach for  $F(T)$ gravity with $f$ essence. We used  the Noether symmetry approach for to determine forms of the physical quantities as $F=C_{1} T^{\frac{3 \left(m-1\right)}{2 \left(n-1\right)}}$, $K = K_{0} \left(Y - \nu u\right)$ and $h = h_{0} u^{m}$ .  Taking the derivative of the scale factor at the time, we can determine the type of the parameter as the Hubble. Next, we have found values for the energy and pressure for the fermion field. Finally, for the our model we obtained is the equation of state parameter as $\omega = \frac{p}{\rho} = - \frac{1}{3} \left(1+2n\right)$. From this equation we can see that is the constant $n$ can take the value $n>1$ then we have $\omega < -1$ that this phase is the phantom phase and if $n=0$ we have $\omega = - \frac{1}{3}$ is the quintessence phase. As was shown earlier for our model, the constant $n\neq1$. However, when $n =1$, for the our model has no physical meaning. Also considered was the case when $n = - \frac{1}{2}$. For this case we have $\rho_{tot} = \frac{4}{3 \left(t-C_{2}\right)^{2}}$ and $p=0$. We can see that this solution for the fermionic field gives us a standart pressureless matter field. 


\end{document}